\documentclass[cameraready]{Interspeech}
\usepackage{dirtytalk}
\usepackage{enumitem}

\title{Lived Experiences of Power and Agency: Achieving Voice Sovereignty in Assistive Speech Technology for Nonbinary Users}

\author[affiliation={1}, orcid=0000-0003-0674-0192, correspondingauthor]{Maxwell}{Hope}
\author[affiliation={2}, orcid=0009-0005-3693-511X]{Juliana}{Francis}
\author[affiliation={2}, orcid=0000-0002-0397-6442]{Joakim}{Gustafson}
\author[affiliation={2}, orcid=0000-0003-1175-840X]{Éva}{Székely}

\address{
    $^1$ University of Delaware, USA \\
    $^2$ KTH Royal Institute of Technology, Sweden
}

\email{maxhope@udel.edu, jfrancis@kth.se, jkgu@kth.se, szekely@kth.se}

\keywords{nonbinary, identity alignment, assistive technology,  agency, human-computer interaction, voice sovereignty}

\usepackage{comment}

\newcommand{\singleindentquote}[1]{%
\vspace{0.15em}
\begin{list}{}{\leftmargin=1em \rightmargin=0em \topsep=0pt \parsep=0pt \itemsep=0pt}
\item[] \textit{``#1''}
\end{list}
\vspace{-0.6em}
}
\usepackage{graphicx}
\usepackage[utf8]{inputenc}
\usepackage[T1]{fontenc}

\begin{document}

\maketitle

\begin{abstract}
Speech-generating devices (SGDs) are essential tools for many speech-disabled individuals, yet current systems often reinforce normative assumptions about identity and communication. Prior research on gender expansive SGD users has largely focused on gendered voice options, with less attention to broader identity alignment and the technological conditions that shape users’ agency. This study examines how gender expansive SGD users experience power over their voices using thematic analysis of qualitative survey data. Three themes emerged: (1) integration of SGDs into daily life, (2) identity expression, and (3) power and agency. Participants expressed the need for more responsive and reliable technology, inclusive and customizable voices, and greater autonomy in defining how their voices sound and are perceived. Ultimately, their perspectives call for a reimagining of SGD systems as tools for reclaiming the power of one’s voice and the ability to define and enact one’s will through it.

\end{abstract}

\let\thefootnote\relax\footnote{Proceedings of Interspeech 2026, 10.21437/Interspeech.2026-2663}

\section{Introduction}
Gender expansive (i.e., transgender and nonbinary) people who rely on speech-generating devices (SGDs) to communicate live at the intersection of two marginalized groups. They are a heterogeneous group themselves, differing in the conditions that bring them to assistive communication, in how they understand and express their gender, and in their linguistic and cultural backgrounds. This makes their lived experience a uniquely valuable  source of insight into assistive voice technology and its design, yet prior work with this community has primarily examined the expression of gender-diverse identity. The present study broadens this scope to what gender expansive SGD users can tell us about technology, voice, and identity more broadly, and about the power and agency they hold over their voices.

\subsection{SGDs and technological integration}
SGDs are a type of high-tech augmentative and alternative communication (AAC) which produce synthetic voices that encode linguistic messages dictated by users through symbols, pictures, or typed messages using a keyboard. For people who use SGDs, the synthetic voice a device produces is closely tied to their identity and to how they are represented to others \cite{wickenden2011whose}. Qualitative research has shown that what users value in these systems extends well beyond message transmission to the contextual factors that shape communication and the personal significance of how they are heard \cite{broomfield2024qualitative}.  Moreover, because these devices are \say{close to the body, inviting frequent interaction} they may become embodied -- that is, in this context, they may become an integral part of one's body \cite{Ihde79-TAP, Nelson19-WTE}. This has also been seen in other assistive devices or tools, such as wheelchairs, where neuroscientific evidence has shown wheelchair users may integrate their wheelchairs into their body schema, becoming a full extension of one's self \cite{pazzaglia2016_embodiment}. 
Because of their close and frequent interactions with these devices, SGD users may indeed embody their devices and the synthetic voice output from them as investigated in \cite{HopeSzekely25_interspeech}. 
If the device is experienced as part of the body, constraints on that device become constraints on the self.

\subsection{Identity and voice}

Voice encodes and indexes many features of identity \cite{Podesva15-VQI, netzorg2025production}, such as gender identity, regional or geographic identity, age, race, educational background, ability or disability. Further, others' speech-based assumptions about an individual's identity reach high agreement between listeners in less than a second \cite{lavan2023time}. Listeners extend the same rapid judgments to synthetic voices, impacting how SGD users themselves are perceived \cite{wickenden2011whose, michel2025s, weinberg2026me}.

Recent AAC-user-led research on culturally and linguistically aligned synthetic voices reports that users want voices reflecting their identity but lack accessible ways to specify how those voices should sound \cite{weinberg2026me}. High-tech AAC systems can support aspects of linguistic identity, including culturally relevant vocabulary and some dialectal pronunciation, but they provide more limited support for morphosyntactic and pragmatic variation (e.g., grammatical patterns, register, and intonation) and frequently require substantial manual programming to implement these features \cite{Frick23-MCA}.
The Mid-Atlantic Gender Expansive Speech (MAGES) corpus \cite{hope_mages}, one of the only functional datasets for training gender expansive synthetic voices, lacks dialectal diversity; therefore, for nonbinary and transgender SGD users outside the Mid-Atlantic United States, it is often insufficient to meet their linguistically diverse needs.

For some nonbinary individuals (i.e., those who do not align exclusively with male or female identity), being unable to accurately convey their gender identity may lead to an increase in misgendering, which may exacerbate mental health issues \cite{Jacobsen23-MHW}. Additionally, regardless of misgendering, everyone deserves a voice that accurately reflects their gender identity. 
Recent work has examined how nonbinary people who use SGDs may accomplish greater voice gender affirmation with their synthetic voices (see \cite{HopeSzekely25_interspeech, Hope23-DSP, SzekelyHope24_interspeech}), highlighting the limitations of
existing SGDs in supporting gender expansive identity expression. These studies, however, focused on whether voices reflected the participant's gender identity, without examining how nonbinary SGD users experience broader identity expression, technological constraints, or their agency and power over their voices. In particular, they did not explore how environmental, technological, and social factors shape users’ ability to enact their identity through their device, nor how users negotiate, resist, or reclaim control over their voices. The present study builds on this previous work by examining SGD users’ experiences more holistically, centering power, agency, and identity beyond gender alone, and identifying the technological and social conditions that facilitate or constrain nonbinary users’ ability to define and enact their voice and convey messages on their own terms.

\subsection{Power, Agency, and Sovereignty}

\emph{Agency} refers to the capability of individuals to act intentionally and make a difference in the social world \cite{Giddens84-COS}, while \emph{power} is the chance that an individual in a social relationship can achieve their own will even against the resistance of others \cite{Weber68-EAS}. Applied to SGDs, agency can be understood as “the ability for users to author and transmit messages enacting their intentions” \cite{Sankaran23-RPU}, and power as the ability to transmit those intentions even when constrained by social expectations. These concepts connect to \emph{sovereignty}, a framework originating in Indigenous thought that centers recognized authority over one’s actions and conditions of participation \cite{Alfred05-WIP}. Recent assistive technology research extends this through \emph{relational sovereignty}, defined as the user’s authority to determine when to act independently or interdependently and to define the goals and terms under which technology is used \cite{Jang26-AST}. From this perspective, \emph{voice sovereignty} for SGD users can be understood as the recognized authority to define, control, and enact one’s voice in ways that reflect one’s identity, intentions, and, when applied relationally, chosen modes of interaction within social contexts.

\subsection{The current study}
AAC users are often left out of AAC research \cite{blasko2025nothing}. As \cite{Corbin25-SIE} states: \textit{\say{[AAC users] contributions to AAC research - via participatory models that position us as equal partners - can, will, and do make that research more meaningful.}} This study invites SGD users to participate in the first stage of participatory action research \cite{Baum06-PAR} by conducting a needs assessment of the community in an open-ended way, letting the community itself dictate what the greatest needs of their community are. This will set the stage for future research to be able to address real, emerging concerns of the community without making assumptions about what the community would like to be addressed. Following the present research, researchers should continue to use a full participatory action research approach inviting SGD users to be part of research teams and involved in the recruitment, data collection, data analysis, and research dissemination. 

\section{Methods and Materials}

\subsection{Participants}
\label{sec:participants}
Three nonbinary SGD users over the age of 18 were recruited from a mailing list of nonbinary SGD users who had previously consented to be contacted for the purposes of research. Participants were compensated 20 USD for their participation. 

\subsubsection{Participant gender and SGD usage}
Participants' gender and SGD usage are shown in Table \ref{tab:participant_gender} and Table \ref{tab:participant_sgd}. Participants were able to select categorical gender in a select-all-that-apply question containing: Man, Woman, Nonbinary, Genderqueer, and Genderfluid. Participants could also enter an open-ended gender description if they wished. 

\begin{table}[th]
  \caption{Participant gender identity information}
  \label{tab:participant_gender}
  \centering
  {\fontsize{8}{10pt}\selectfont
  \begin{tabular}{l p{0.50\columnwidth} p{0.30\columnwidth}}
    \toprule
    \textbf{ID} & \textbf{Categorical gender} & \textbf{Open-ended gender} \\
    \midrule
    P1 & Man, Nonbinary, Genderqueer, Genderfluid & --- \\
    P2 & Nonbinary, Genderqueer & Transmasc \\
    P3 & Nonbinary, Genderqueer & Neuroqueer \\
    \bottomrule
  \end{tabular}
  }
\end{table}

\begin{table}[th]
  \caption{Participant SGD usage}
  \label{tab:participant_sgd}
  \centering
  {\fontsize{8}{10pt}\selectfont
  \begin{tabular}{l p{0.4\columnwidth} l}
    \toprule
    \textbf{ID} & \textbf{Type of SGD} & \textbf{Frequency of SGD use} \\
    \midrule
    P1 & iPad with multiple apps and Accent with LAMP and Unity & Every day \\
    P2 & iPad Mini with TD Snap (used only for AAC) & Most days \\
    P3 & Grid Pad 13 and Grid for iPad on iPad & Every day \\
    \bottomrule
  \end{tabular}
  }
\end{table}

\subsection{Survey}

Participants completed two anonymous surveys one week apart via Google Forms. The initial survey asked for information concerning each participant's gender identity and SGD usage (see section \ref{sec:participants} above) and then presented a series of questions asking them to describe their daily experiences with their SGD and how they can (or cannot) express their identity through their SGDs. At the conclusion of the first survey, participants were instructed to take a week to reflect on the questions as they went about their daily life prior to completing the second survey. The second survey was provided to participants one week later, and included all questions except the demographic questions; this was so that after a week of reflection, participants could add to their responses in case of any new insights. 
\subsubsection{Survey questions}
Listed below are the open-ended survey questions we asked participants on both surveys. 

\begin{enumerate}[label=\textbf{Q\arabic*}, leftmargin=2.2em, itemsep=1pt, topsep=2pt]
  \item Describe an average day from the moment you get up to the time you go to sleep, focusing on how you use your SGD.
    
    \item Describe situations that you find challenging when using your SGD.
    
    \item Are there times of day or types of situations where using your SGD feels easier or harder?
    
    \item Do you prepare or rehearse things you want to say using your SGD in advance? Why or why not?
    
    \item Does using your SGD limit your ability to express your identity? If yes, please describe a time where this occurred.
    
    \item How could your SGD device be improved to allow you to express your identity more effectively?
    
    \item How well does the voice(s) you use on your SGD reflect your identity?
    
    \item How would you customize your voice(s) on your SGD to better fit your identity, if at all?
\end{enumerate}

\subsection{Analysis}

We used an inductive thematic analysis approach \cite{BraunClarke06-UTA} conducted on responses to the semi-structured theoretically informed survey questions. As a result, major themes identified are heavily influenced by the survey questions themselves. To look deeper into the specifics, we further divided themes identified into several subthemes where applicable. Quotes from subthemes are further divided into facilitators and barriers where applicable.

\section{Results}
Three major themes emerged from the data: integration of SGDs into daily life, identity expression, and a cross-cutting theme of power and agency. Selected quotes are presented in each theme below. Full counts in each subtheme are shown in Figure 1.

\begin{figure}[t]
\centering
\includegraphics[width=1.0\columnwidth]{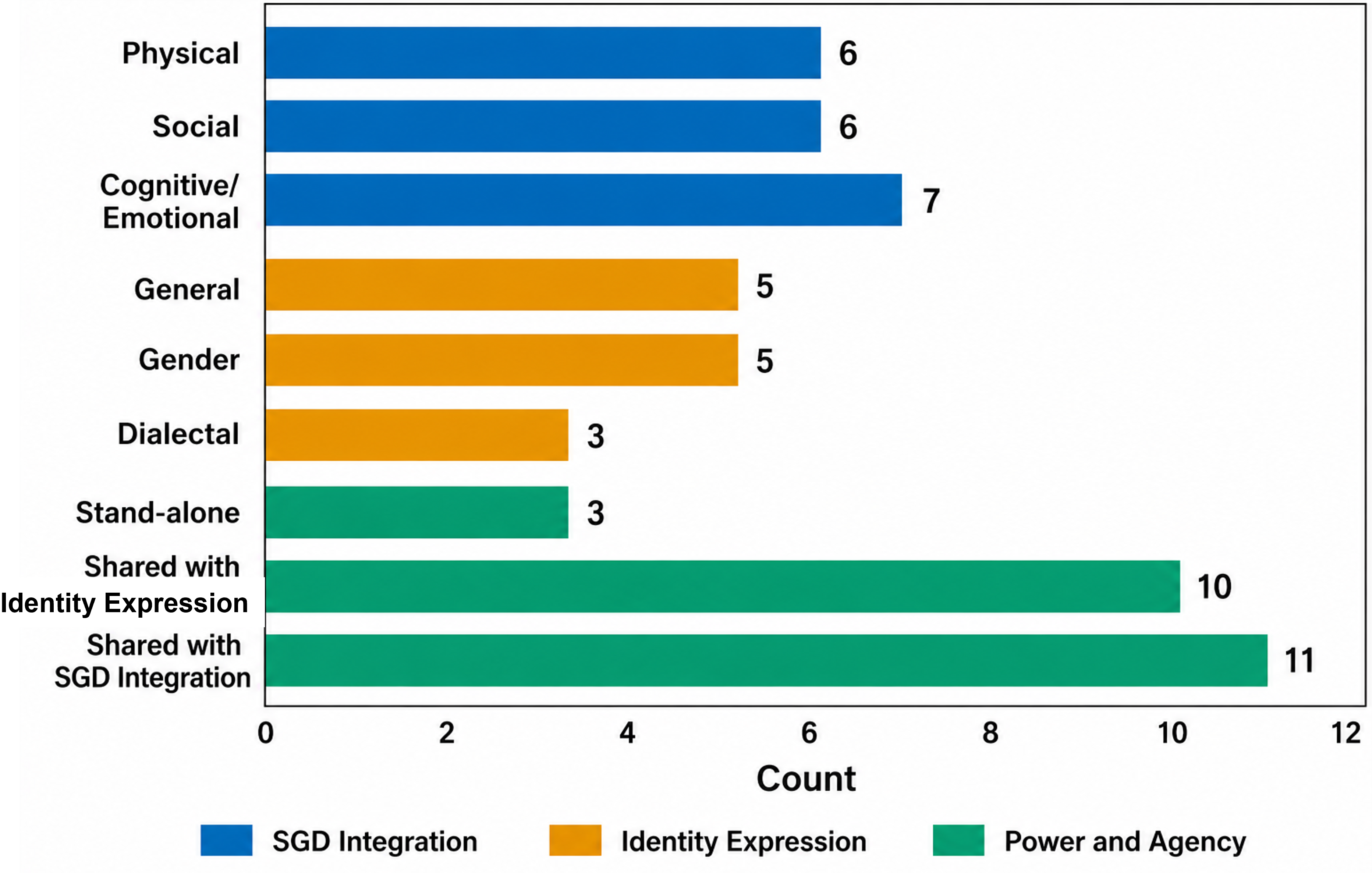}
\caption{Bar graph showing the number of quotes in each theme (SGD Integration = blue, Identity Expression = yellow, Power and Agency = green) broken down by subtheme; Power and Agency, as a cross-cutting theme, contains stand-alone quotes as well as quotes shared with Identity Expression and SGD Integration.}
\label{fig:sgd_diagram}
\end{figure}

\subsection{Integration of SGDs into daily life}
\subsubsection{Physical integration}
Participants reported that they kept their devices close to them and frequently interacted with them, suggesting that these devices become integrated into the concrete and physical realities of day-to-day life. 

\noindent\textbf{Facilitators of physical integration:}

\singleindentquote{My iPad stays on me or nearby and I use it to communicate.}
\vspace{0.25em}

\singleindentquote{Put on device (worn on strap)\ldots{} Device used as needed throughout [the day].}
\vspace{0.25em}

\singleindentquote{My communication device or phone generally come with me unless they are charging.}
\vspace{0.25em}

\singleindentquote{[It’s easier to use my SGD] laying on my bed with my device mounted.}
\vspace{0.5em}

\noindent\textbf{Barriers to physical integration:}

\singleindentquote{Loud places.}
\vspace{0.25em}

\singleindentquote{Anywhere where it’s too loud, or wet.}
\vspace{0.25em}

\subsubsection{Cognitive and emotional integration}
Participants described cognitive and emotional factors that facilitate or hinder communication in daily life.

\noindent\textbf{Facilitators of cognitive and emotional integration:}
\singleindentquote{I use [my SGD] most often in the evenings, when my energy is lower.}
\vspace{0.25em}

\singleindentquote{[Scripting is] really important cognitively for me.}
\vspace{0.25em}

\singleindentquote{Sometimes my communication device comes with me mostly for comfort.}
\vspace{0.25em}

\noindent\textbf{Barriers to cognitive and emotional integration:}
\singleindentquote{Extreme fatigue\ldots{} makes it harder.}
\vspace{0.25em}

\singleindentquote{Slow if I don’t [script].}
\vspace{0.25em}

\singleindentquote{Trying to answer questions\ldots{} I end up really struggling.}
\vspace{0.25em}
\singleindentquote{I have a lot of trauma associated with trying to force speech and that makes it hard to use my device even though I need to.}
\vspace{0.5em}

\subsubsection{Social integration}
Participants described that working with friends and those who are understanding facilitates SGD integration while it was harder with people who are not understanding. 

\noindent\textbf{Facilitators of social integration:}
\singleindentquote{[Using my SGD is] easier with partner or friends.}
\vspace{0.25em}

\singleindentquote{When interacting with people who understand, [using my SGD is] easier.}
\vspace{0.25em}

\noindent\textbf{Barriers to social integration:}
\singleindentquote{People who don’t wait.}
\vspace{0.25em}

\singleindentquote{Anything with speed being relevant. People try to rush me.}
\vspace{0.25em}

\singleindentquote{Harder with family.}
\vspace{0.25em}

\singleindentquote{People who don’t understand/accept AAC.}
\vspace{0.25em}

\subsection{Identity expression}
\subsubsection{General/broad identity expression}
\noindent\textbf{Facilitators of general identity expression:}
Participants described their SGD as enabling identity expression that was not possible through speech alone.

\singleindentquote{It usually helps me express my identity.}
\vspace{0.25em}

\singleindentquote{Better than my mouth voice.}
\vspace{0.25em}

\noindent\textbf{Barriers to general identity expression:}
Participants had to sacrifice their expression of identity to sound intelligible. 

\singleindentquote{The voice I use doesn't always feel like the best fit.}
\vspace{0.25em}

\singleindentquote{Better than my mouth voice and still badly.}
\vspace{0.25em}
\singleindentquote{I chose a voice based on intelligibility and not identity.}
\vspace{0.25em}

\subsubsection{Gender identity expression}
\noindent\textbf{Facilitators of gender expression:}
Participants expressed a desire for voices aligned with their gender identity, particularly voices outside of binary gender norms.

\singleindentquote{More gender neutral and queer voices.}
\vspace{0.25em}

\singleindentquote{More gender options.}
\vspace{0.25em}

\singleindentquote{Have something which people can't immediately identify as Male or Female.}
\vspace{0.25em}

\singleindentquote{Make it gender weird.}
\vspace{0.5em}
\noindent\textbf{Barriers to gender expression:}
Participants also described technological limitations that prevented alignment between their SGD voice and gender identity.

\singleindentquote{I use the Acapella Scott voice pitched down, and TD Snap didn’t used to allow changing pitch so I had to use a voice that fit worse.}
\vspace{0.5em}

\subsubsection{Dialectal/geographic identity expression}
\noindent\textbf{Facilitators of dialectal expression:}
Participants wanted voices reflecting their regional and geographic identities. While these are largely unavailable right now, participants imagined them as good facilitators to express this facet of their identity.

\singleindentquote{Appalachian accent please.}
\vspace{0.25em}

\singleindentquote{Better match my location.}
\vspace{0.25em}

\singleindentquote{Gender neutral new england american english voice.}
\vspace{0.25em}

\subsection{Power and agency}

This major theme cuts across the other two themes and ties them together. For both integration of SGDs into daily life and identity expression, the ability to enact one's will despite others' resistance frequently comes up for participants. 

\noindent\textbf{Facilitators of power and agency:}
Participants described reclaiming ownership, autonomy, and control over their voice through customizing their SGD. They intentionally shaped their voice as an act of agency rather than simply accepting available options. Technology could support this agency by enabling greater flexibility, customization, and choice.

\singleindentquote{Its my voice and i want the power.}
\vspace{0.25em}

\singleindentquote{I want something that is uncanny valley purposefully.}
\vspace{0.25em}

\singleindentquote{Make it easier for me to have multiple options to swap between moment to moment.}
\vspace{0.5em}

\noindent\textbf{Barriers to power and agency:}
Participants described ways in which technological limitations constrained their ability to act with autonomy. Social expectations of conversational speed also constrained participants' agency. Safety further influenced one participant's ability to enact their will using their voice.

\singleindentquote{The inability to move quickly and the logistics [...] end up limiting me.} 
\vspace{0.25em}

\singleindentquote{People who don’t wait [make it challenging to use my SGD].}
\vspace{0.25em}

\singleindentquote{People try to rush me, that makes me struggle.}
\vspace{0.25em}

\singleindentquote{Sometimes [it’s challenging to use my SGD] if I feel unsafe.}
\vspace{0.25em}
\section{Discussion}

This study presents an in-depth investigation into the current needs of three nonbinary SGD users. While the sample is small, the qualitative approach yields richer data, appropriate for an investigation of a niche, under-researched community \cite{Corbin25-SIE}. 

\subsection{Facilitators and barriers of power in SGDs}
We found evidence that SGDs facilitate the embodiment of power in expanding the daily experiences and identity expression of nonbinary SGD users within this study. Participants were thoughtful about what facilitates their lived experiences and identity expression, while considering possibilities of what could enhance this if they were given more power and agency, and if those developing SGDs listened to their input in the future. These current and hoped-for facilitators include faster and more responsive SGDs, more gender options (including ones subverting binary norms), and more dialectal options. 

We found several areas where barriers to enacting one's will and expressing identity exist. Participants described prioritizing intelligibility over identity when selecting SGD voices, even when those voices did not reflect their sense of self. One participant explained that they \say{chose a voice based on intelligibility and not identity}, illustrating how communicative effectiveness often takes precedence over personal authenticity. This echoes phenomenological accounts such as \cite{Howery18-WMV}, where users revert to more generic or standardized voices because they are more readily understood, even when these feel less personal or identity-aligned. Many barriers also occurred in social interactions, reflecting limited relational sovereignty; participants often felt rushed and reported that others would not wait for them to communicate.

\subsection{Implications for future research and SGD design}
Recent advances in text-to-speech (TTS) architectures show that many features requested by our participants are now technically feasible. 
Systems including Parler-TTS \cite{lyth2024natural}, StyleTTS-ZS \cite{li2025styletts}, VoiceCraft \cite{peng2024voicecraft}, CosyVoice 3 \cite{du2025cosyvoice}, and IndexTTS2 \cite{zhou2026indextts2} demonstrate complementary forms of control, including natural-language prompting of speaker identity, style, accent, and prosody; zero-shot speech editing; multilingual and dialectal voice cloning in realistic speech conditions; pronunciation inpainting; and duration/timing control with disentangled control of timbre and expressive style. Speaker interpolation methods \cite{francis2025interpolating} suggest ways users might explore continuous regions of voice space rather than selecting only from fixed, discrete voice categories.
These capabilities map directly to needs identified in our study, including identity-aligned voices beyond binary gender categories, regional accents, flexibility in voice customization, and conversational timing that supports turn-taking.

While state-of-the-art TTS systems can provide the technical capabilities needed for identity-aligned voices, they do not yet operationalize identity-centered design in assistive contexts. Control mechanisms often rely on stereotyped or gendered descriptors, dialect layers are not always independent of timbre, and evaluation metrics remain focused on intelligibility and similarity rather than ownership, identity alignment, or perceived agency. In addition, practical concerns such as latency, privacy, licensing feasibility, and consent workflows remain underdeveloped in SGD deployment. 
These stakes grow sharper as SGDs begin to incorporate LLMs that generate or complete utterances: the more a device contributes to what is said, the more critical it becomes to preserve the user's agency and authorship over their voice \cite{griffiths2025ai}.

Future work should center the active involvement of SGD users in the design and evaluation of communication technologies to support voice sovereignty \cite{blasko2025nothing}. This may include co-design sessions with SGD users to prototype voices that reflect gender expansive identities and regionally specific linguistic features; collaborative exploration of interactional norms (e.g., training conversation partners to wait); and designing interfaces that visually signal when a user is composing a message.

\section{Conclusion}
In \cite{Howery18-WMV}, the author describes how devices running out of battery can cause an instant and abrupt stop to the embodiment of the device: \textit{\say{Any chance of continuing to meaningfully engage in embodiment or hermeneutic relation with the device has been abruptly interrupted by the device itself. It has become something other than the device that allows one to speak but rather an other that demands I attend to its needs—in this case the need for power.}} Humans, too, have a need for power. It is up to the linguists, speech technologists, speech scientists, and engineers working on SGD research and development to finally, and at once, recharge their work by ceding power to members of the SGD community, especially those who are the most marginalized within that community.

\section{Acknowledgments}
The authors thank the gender expansive participants who took part in this study. Despite the many challenges currently facing transgender and nonbinary people around the world, you are finding ways to reclaim your power. 
This work is funded in part by the WASP funded project Real-time context-aware speech prosthesis for conversational interaction.

\section{Generative AI Use Disclosure}
AI tools were used to assist with table coding and with polishing portions of the text; all AI-assisted content was reviewed, verified, and modified by the authors.

\bibliographystyle{IEEEtran}
\bibliography{mybib}

\end{document}